\documentclass[aps,prd,reprint,twocolumn,superscriptaddress,longbibliography,nofootinbib,floatfix,showpacs]{revtex4-1}
\usepackage{epsfig}
\usepackage{amsmath,amssymb,amsfonts}
\usepackage{hyperref}
\usepackage{mathrsfs}
\usepackage{bbm}
\usepackage{slashed}
\usepackage{graphicx}
\usepackage{verbatim}
\usepackage[usenames]{color}

\newcommand{\be}{\begin{equation}}
\newcommand{\ee}{\end{equation}}
\newcommand{\bi}{\begin{itemize}}
\newcommand{\ei}{\end{itemize}}
\newcommand{\bea}{\begin{eqnarray}}
\newcommand{\eea}{\end{eqnarray}}
\newcommand{\bracket}[2]{\bra{#1}\,#2\rangle} 
\newcommand{\bra}[1]{\langle\,#1\,|}          
\newcommand{\ud}{\mathrm{d}}

\newcommand{\LCm}{{\scriptscriptstyle -}} 
\newcommand{\LCp}{{\scriptscriptstyle +}}
\newcommand{\LCpm}{{\scriptscriptstyle \pm}}

\newcommand{\LCperp}{{\scriptscriptstyle \perp}}

\begin{document}

\title{Localisation in worldline pair production and lightfront zero-modes}
\author{Anton Ilderton}
\email[]{anton.ilderton@chalmers.se}
\affiliation{Department of Applied Physics, Chalmers University of Technology, SE-41296 Gothenburg, Sweden}

\begin{abstract}
The nonperturbative probability of pair production in electric fields depending on lightfront time is given exactly by the locally constant approximation. We explain this by showing that the worldline path integral defining the effective action contains a constraint, which localises contributing paths on hypersurfaces of constant lightfront time. These paths are lightfront zero-modes and there can be no pair production without them; the effective action vanishes if they are projected out.
\end{abstract}

%
\maketitle

\section{Introduction}
The possibility of spontaneous pair production in external fields is one of the most well-known nonperturbative results in quantum field theory, with the example of a constant and homogeneous electric field detailed in~\cite{Sauter:1931zz,Heisenberg:1935qt,Schwinger:1951nm}. Going beyond constant fields leads to richer mathematics and physics, with time-dependent electric fields $E(x^0)$ being the case most commonly studied~\cite{Hebenstreit:2008ae,Hebenstreit:2009km,Dumlu:2011rr,Dumlu:2013pma}. There are though only a few examples of specific field shapes in which the pair production probability can be calculated exactly. To instead estimate the probability in a given field, describing e.g.\ focussed laser pulses~\cite{Bulanov:2004de,Dunne:2009gi,Schutzhold:2008pz,DiPiazza:2009py,Bulanov:2010ei,Gonoskov:2013ada}, the locally constant approximation (LCA) is often used. In this approximation, the constant electric field $E$ and a volume factor appearing in the famous expressions due to Heisenberg and Euler~\cite{Heisenberg:1935qt}, and Schwinger~\cite{Schwinger:1951nm}, are replaced with $E(x)$ and a integral over~$x^\mu$, respectively. Exactly solvable cases show, though, that the LCA is not exact in general~\cite{NikishovSimplest,Dunne:2004nc,Gavrilov:2009vf}.

Pair production in longitudinal electric fields $E^3(x^\LCp)$, depending on lightfront time $x^\LCp=x^0+x^3$~\cite{Dirac:1949cp}, was investigated in~\cite{Tomaras:2001vs,Tomaras:2000ag,Woodard:2001ai,Woodard:2001hi} using lightfront quantisation, i.e.\ quantisation on hypersurfaces of constant $x^\LCp$~\cite{Brodsky:1997de,Heinzl:2000ht}. In this approach, a prescription is required for dealing with zero-modes, states for which the momentum $p^\LCp$ vanishes, due to ubiquitous $1/p^\LCp$ terms. It was shown in~\cite{Tomaras:2001vs,Tomaras:2000ag} that standard regularisations of zero-modes would imply that there is no pair production in fields $E^3(x^\LCp)$, contradicting the constant field results of~\cite{Heisenberg:1935qt,Schwinger:1951nm}. This problem was solved in~\cite{Tomaras:2001vs,Tomaras:2000ag} by quantising on two lightlike hypersurfaces, one of fixed $x^\LCp$ and one of fixed $x^\LCm=x^0-x^3$~\cite{Bogolyubov:1990kw,McCartor:1996nj}. Although effectively abandoning the usual lightfront approach, this allowed control of the zero-modes and lead to a nonzero pair production probability (which was also recovered from a functional computation of the effective action in~\cite{Fried:2001ur}). Surprisingly, the probability was found to be given {\it exactly} by the LCA, and for {\it arbitrary} field dependence on $x^\LCp$.

Explaining why this remarkably simple result should hold in $x^\LCp$--dependent fields is the goal of this paper. As it has been claimed that the LCA applies to $x^0$--dependent fields, despite known counterexamples, it is also important to understand the difference between these two cases. Further, a better understanding of how spatial, temporal~\cite{Dunne:2005sx}, and here lightlike inhomogeneities affect pair production will be useful in the search for analytic results in ever more complex field configurations~\cite{Schneider:2014mla}. An additional motivation is to cement the connection between nonperturbative pair production (one of the most exciting physical phenomena in strong fields~\cite{Dunne:2008kc}) and zero-modes (one of the oldest theoretical issues in lightfront field theory)~\cite{Ji:1995ft}.


Here we will reconsider pair production in fields $E^3(x^\LCp)$ using the worldline formalism, reviewed in~\cite{Schubert:1996jj}, which has found particular success in application to pair production in external fields~\cite{Dunne:2005sx,Dunne:2006st}. The paper is organised as follows. In Sect.~\ref{SECT:EA} we compute the one-loop effective action using the worldline path integral, giving a simple derivation of the result in~\cite{Tomaras:2001vs,Tomaras:2000ag,Fried:2001ur}. We will see explicitly why the effective action localises, and what role the zero-modes play in this. In Sect.~\ref{SECT:ZM} we give extensions of our results, compare with pair production in time-dependent fields $E(x^0)$, for which the LCA does  not hold, and relate the zero-mode contributions to the triviality of the lightfront vacuum. We conclude in Sect.~\ref{SECT:C}.

\section{The effective action}\label{SECT:EA}
The probability of pair production in an external field and the effective action $\Gamma$ are related by $\mathbb{P}_\text{pairs}=1-e^{-2\text{Im }\Gamma}$. The one-loop effective action is given by the functional integral
\be\label{Gamma}
	\Gamma = \int\limits_\delta^\infty\!\frac{\ud T}{T}\int\!\mathcal{D}x^\mu \ \exp i S[x;T] \;,
\ee
where $\delta$ is a UV cutoff and the worldline particle action $S$ is a function of the {\it closed} path $x^\mu$ and the intrinsic length of the path, $T$,
\be
	S[x;T] = -\frac{m^2T}{2} - \int\limits_0^1\!\ud\xi T \bigg[\frac{{\dot x}^2}{2T^2}  + e\frac{{\dot x}^\mu}{T} A_\mu(x) \bigg]\;.
\ee
(For clarity we consider scalar particles; the extension to spinors is given below.) A closed path $x^\mu$ can be decomposed into a constant, average, piece $x_c^\mu$ and an orthogonal oscillatory piece $y^\mu$. See Appendix~\ref{APPA} for details; here we need only that the measure on these variables is, for each coordinate~\cite{Polyakov:1987ez},
\be\label{Measure1}
	\int\!\mathcal{D} x =	 \sqrt{T}\int\!\ud x_c \int\!\mathcal{D}y.
\ee
We will ignore constant prefactors, which can be recovered by comparison with Schwinger's result~\cite{Schwinger:1951nm}. Lightfront coordinates are defined by $x^\LCpm = x^0\pm x^3$, and the transverse directions are $x^\LCperp=\{x^1,x^2\}$. Following~\cite{Tomaras:2001vs,Tomaras:2000ag} we take $A_\LCm\equiv A_\LCm(x^\LCp)$ and all other components zero ($\Gamma$ is trivially gauge invariant). Writing $a := -eA_\LCm$, the longitudinal electric field\footnote{Our chosen field does not satisfy Maxwell's equations in vacuum, but nor do time-dependent fields $E(x^0)$ normally considered in the worldline approach. The phenomenological motivation for studying the latter class is that they model (neglecting spatial structure) the standing wave formed in the focus of two colliding laser pulses. Similarly, our field may be taken to model (neglecting spatial structure) the longitudinal electric field of a strongly {\it focussed} laser pulse, see~\cite{Davis:1979zz,McDonald:1995} for the form of such fields.} is
\be
	eE^3(x^\LCp) = 2 a'(x^\LCp) \;.
\ee
Extensions will be discussed below. Note that $a'=$ constant is the constant field case~\cite{Schwinger:1951nm}. We now evaluate the effective action.

\subsection{Localisation and zero-modes}
Lightfront zero-modes correspond to vanishing momentum $p^\LCp$, i.e.\ dynamics within a lightlike hypersurface $x^\LCp=$ constant. We are therefore particularly interested in contributions to (\ref{Gamma}) from paths which lie in such hypersurfaces. In lightfront quantisation, some degrees of freedom are typically constrained, and it is in solving these constraints that the issue of zero-modes must be confronted, for example in defining the operator $1/p^\LCp$~\cite{Brodsky:1997de,Heinzl:2000ht}. Something similar will happen here: we will first integrate out the longitudinal degrees of freedom $x^\LCm$, which will generate a constraint leading us to the zero-modes.

The part of the action depending on $x^\LCm$ is
\be\label{GP}
	S' := -\int\limits_0^1\!\ud\xi\, T\ \bigg[\frac{{\dot y}^\LCm{\dot y}^\LCp}{2T^2} - \frac{{\dot y}^\LCm}{T} a(x_c^\LCp + y^\LCp) \bigg]\;.
\ee
Since $x_c^\LCm$ does not appear, the $x^\LCm_c$--integral contributes only a volume $V^\LCm$. The boundary conditions on the paths allow us to integrate by parts in (\ref{GP}) and remove all derivatives from $y^\LCm$. The $y^\LCm$ integral can then be performed and gives a delta functional on the space of paths $y^\LCp$:
\be\label{Constraint}
	\int\!\mathcal{D}y^\LCm e^{iS'} = \delta\big[\tfrac{1}{2T^2}\ddot{y}^\LCp - \tfrac{1}{T}\dot{a}(x_c^\LCp+y^\LCp)\big] =: \delta[O[y^\LCp]] \;.
\ee
Thus, the paths in $\{x^\LCp,x^\LCperp\}$--space which can contribute to the effective action are constrained by (\ref{Constraint}), which we must solve in order to proceed. Although the constraint looks highly nonlinear our key result is that, because of the boundary conditions, the delta functional has support only on $y^\LCp\equiv0$, as we now show. Integrating the equation $O[y^\LCp]=0$, implied by (\ref{Constraint}), we obtain
\be\label{Constraint1-5}
	{\dot{y}}^\LCp(\xi)\big|_{\xi_1}^{\xi_2} = 2T a(x_c^\LCp + y^\LCp(\xi))\big|_{\xi_1}^{\xi_2} \;.
\ee
which simply confirms that $\dot y$ obeys the same boundary conditions as $y$. If we instead expand $\dot{a} = {\dot y}^\LCp a'$ in (\ref{Constraint}) and use an integrating factor, we obtain the implicit relation
\begin{figure}[t!]
	\centering\includegraphics[width=0.8\columnwidth]{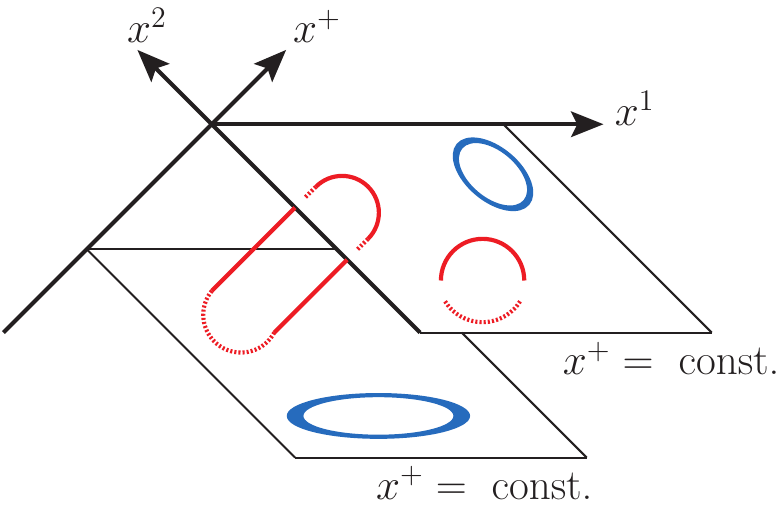}
	\caption{\label{FIG:BIDRAG} Contributing paths (after integrating out $x^\LCm$): the blue loops, which lie within hypersurfaces of constant $x^\LCp$, contribute to the effective action. The red loops, which cut those hypersurfaces, do not contribute.}
\end{figure}
\be\label{Constraint2}
	\dot{y}^\LCp(\xi)\exp\bigg[-2T\!\int\limits_{\xi_0}^{\xi}\ud \tau\ a'(x^\LCp_c + y^\LCp (\tau))\bigg] = \dot{y}^\LCp(\xi_0) \;.
\ee
Assume first that $E^3 = 2a'>0$. Then the exponent in~(\ref{Constraint2}) is strictly negative, and it is impossible to satisfy the boundary conditions unless $\dot{y}(0)=\dot{y}(1)=0$. Then (\ref{Constraint2}) implies that $\dot{y}(\xi)\equiv 0$. It follows from (\ref{Constraint1-5}) that the purely oscillatory $y(\xi)$ is constant, and the only allowed constant is $y(\xi)\equiv 0$.  Thus the delta functional in (\ref{Constraint}) has support {\it only} on $y^\LCp=0$, and we have
\be\label{delta}
	\int\!\mathcal{D}y^\LCp \delta[O] = \int\!\mathcal{D}y^\LCp \delta\big[y^\LCp\big] \text{Det}^{-1}\frac{\delta O}{\delta y^\LCp} = \text{Det}^{-1}O'\;,
\ee
where, from here on dropping the $c$--subscript from $x^\LCp_c$,
\be
	O' = \frac{\delta O[y^\LCp]}{\delta y^\LCp}\bigg|_{y^\LCp=0} = \bigg[\frac{1}{2T^2}\frac{\ud^2}{\ud \xi^2} - \frac{a'(x^\LCp)}{T}\frac{\ud}{\ud\xi}\bigg]\delta(\xi-\xi') \;.
\ee
Having solved the constraint (\ref{Constraint}), we see that the effective action (in the considered class of fields) is supported entirely on loops for which $y^\LCp=0 \iff \dot{x}^\LCp=0$. In other words, the effective action is supported entirely on loops lying in hyperplanes of constant $x^\LCp$, which are the lightfront zero-modes. See Fig.~\ref{FIG:BIDRAG}. This is what localises the effective action, as the constraint turns the functional integral $\mathcal{D}x^\LCp$ into an ordinary sum $\ud x^\LCp$ over contributions $\text{Det}^{-1}O'$ from hypersufraces of fixed lightfront time $x^\LCp$:
\be
	\int\!\mathcal{D}x^\LCp\ \delta\big[O[y^\LCp]\big] = \int\!\ud x^\LCp\ \text{Det}^{-1}O'  \;.
\ee
Since the field has only a one-dimensional homogeneity, and because contributing paths are localised in this dimension, (\ref{Gamma}) will become equal to that obtained by applying the LCA to the constant field result. We now make this explicit by  evaluating the determinant in (\ref{delta}) and hence the effective action. For the case that $E^3$ can change sign, see Appendix~\ref{SECT:TECKEN}.

\subsection{Evaluation}
The remaining transverse integrals in (\ref{Gamma}) are those of the free theory, and contribute the standard factor $V^\LCperp/T$. The effective action then takes the form
\be\begin{split}\label{GP0}
	\Gamma 	&= V^\LCperp V^\LCm \int\limits_\delta^\infty\!\frac{\ud T}{T} e^{-im^2T/2}\int\!\ud x^\LCp\ \text{Det}^{-1}O'\;.
\end{split}
\ee
The determinant can be evaluated by standard means, or by returning to (\ref{GP}), replacing $a\to a'(x^\LCp_c)\dot{y}^\LCp$, and performing the resulting elementary integrals as in Appendix~\ref{APPA}. The result is
\be\label{DETRES}
	\text{Det }O'= \prod\limits_{n=1}^\infty \frac{1}{T^4}\bigg( 1 + \frac{T^2a^{\prime2}(x^\LCp)}{n^2\pi^2}\bigg)\to T^2\frac{\sinh Ta^{\prime}(x^\LCp)}{T a^{\prime}(x^\LCp)}\;, 
\ee
using zeta-function regularisation, in the second step, to define the infinite product. Inserting the inverse of (\ref{DETRES}) into (\ref{GP0}) we arrive at
\be\label{Gamma2}
	\Gamma= V^\LCperp V^\LCm\!\!\int\!\ud x^\LCp\! \int\limits_\delta^\infty\!\frac{\ud T}{T^2} e^{-im^2T/2} \frac{a^{\prime}(x^\LCp)}{\sinh T a^{\prime}(x^\LCp)}\;.
\ee
The integral is convergent in the IR but divergent in the UV, $T\to0$. Subtracting the two problematic terms~\cite{Schwinger:1951nm}, and changing variables $T\to 2T/m^2$, leaves
\be\label{Gamma3}
	\Gamma= V^\LCperp V^\LCm\!\!\int\!\ud x^\LCp\! \int\limits_0^\infty\!\frac{\ud T}{T^2} e^{-iT} \bigg[\frac{\varepsilon(x^\LCp)}{\sinh T \varepsilon(x^\LCp)} - \frac{1}{T}+ \frac{T\varepsilon^2(x^\LCp)}{6}\bigg]\;,
\ee
where $\varepsilon(x^\LCp)=eE(x^\LCp)/m^2 = E(x^\LCp)/E_S$. This exact result agrees with that given by the LCA applied to the constant field case; $E$ is replaced by $E(x^\LCp)$ and a nontrivial integral $\ud x^\LCp$ appears instead of an infinite volume $V^\LCp$. Thus we neatly recover (the scalar version of) the results in~\cite{Tomaras:2001vs,Tomaras:2000ag,Fried:2001ur} from the worldline formalism. 

As a function of $T$, the integrand in (\ref{Gamma3}) has no UV or IR divergences, or poles on the real line. It has an imaginary part only because of the imaginary exponent. To evaluate $\text{Im }\Gamma$ we rotate the $T$-contour clockwise to, but not past, the imaginary axis, where the function does have poles. The contour then runs down the imaginary axis, with semicircular deviations into the lower-right quadrant~\cite{Holstein:1999ta}. The on-axis segments contribute to $\text{Re } \Gamma$, whereas the semicircles contribute to $\text{Im }\Gamma$, the final result for which is
\be\label{Gamma4}
	\text{Im }\Gamma = V^\LCperp V^\LCm \int\!\ud x^\LCp \sum\limits_{n=1}^\infty \frac{(-1)^{n+1}}{n^2} \varepsilon^2(x^\LCp) e^{-n\pi/\varepsilon(x^\LCp)} \;.
\ee
The extension to QED proper is direct. The spin factor to be inserted into the integrand of~(\ref{Gamma}) is~\cite{Polyakov:1987ez,Schubert:1996jj}
\be
	\int\!\mathcal{D}\psi\ \exp i \int\!\ud\xi T \bigg[\psi^\mu i {\dot\psi}^\nu + ie \psi^\mu F_{\mu\nu}(x)\psi^\nu\bigg] \;.
\ee
For our choice of field, the $x^\LCm$--integral can be performed as above, which localises the rest of the integrand, including the spin factor. The fermion integrals can then be performed as in the constant field case, and the final result is still given by the LCA.

\section{Discussion}\label{SECT:ZM}
 %
\subsection{Comparison with $E(x^0)$ and Euclidean space methods}
Here we discuss the similarities and differences between pair production in fields $E^3(x^\LCp)$, for which the LCA holds, and $E^3(x^0)$, for which it does not. Since worldine integrals are often computed in Euclidean space, where one can take advantage of the fact that instanton contributions dominate~\cite{Dunne:2005sx,Dunne:2006st}, as well as employ numerical methods~\cite{Gies:2001tj,Gies:2005bz}, we will also compare Minkowski and Euclidean approaches.

We begin with the constant field case, and consider two methods of computing the effective action. The first is to rotate to Euclidean space and calculate e.g.\ the dominant instanton contributions. Alternatively, one can compute the effective action as we did above, and find that paths are localised in hypersurfaces $x^\LCp=$ constant. Either method recovers the result of~\cite{Heisenberg:1935qt,Schwinger:1951nm}. (Since the constant field strength only singles out the $z$--direction, it is easy to check that one can equally well perform the $x^\LCp$ integral first, to localise the paths in $x^\LCm$. When the longitudinal field also depends {\it nontrivially} on $x^\LCp$, though, it is this direction which is singled out and in which localisation occurs.)

The connection between these two approaches can be seen through (\ref{Constraint2}) in the constant field case. The equation has no nontrivial solutions in Minkowski space, but rotating to Euclidean $T$-space, the boundary conditions can be satisfied if 
\be\label{periodiskt}
	T = \frac{2n\pi}{eE} \;, \quad n\in\mathbb{Z} \;.
\ee
This is both the periodicity condition for instanton solutions, which give the dominant contribution to the effective action, and the locations of the poles in (\ref{Gamma3}) after rotating to Euclidean $T$-space~\cite{CS}. 

Now let the field depend nontrivially on $x^\LCp$, as we have studied. There are again two methods -- in Minkowski space, we proceed as above, with the delta-functional being key to obtaining the effective action. Or, we can rotate to Euclidean space and investigate the instanton structure. This is likely to be more challenging than in the case of a constant field; note for example that the $x^\LCp$--dependent field becomes a complex function after the rotation of $x^0$ to Euclidean time, $x^0\to -i x^4$,
\be
	E^3(x^\LCp) =E^3(x^0 + x^3)  \overset{\text{rotate}}{\longrightarrow} E^3(-ix^4+x^3) \;,
\ee
so that the roles of both complex and real instantons must be considered~\cite{Dumlu:2011cc}. Further, it seems unlikely that the constraint (\ref{Constraint}) will appear directly in Euclidean space, even though (\ref{Constraint2}) suggests that a periodicity similar to (\ref{periodiskt}) should still be of relevance. It is though not at all apparent how to relate this to the instanton or pole structure~\cite{CSP}. Therefore, it is an open question as to precisely which mechanism localises the effective action in the Euclidean calculation -- this will be considered elsewhere.

We now come to the case of a time--dependent field $E^3(x^0)$. The Euclidean space calculation is well-known. Let us try the Minkowski approach, as above. A potential for $E^3(x^0)$ is given simply by replacing $A_\LCm(x^\LCm)\to A_\LCm(x^0)$ and, recalling $x^0=(x^\LCp+x^\LCm)/2$, the relevant terms of the worldline action become
\be\label{nonlinear}
	\int\limits_0^1\! \frac{{\dot x}^\LCm{\dot x}^\LCp}{2T} + {\dot x}^\LCm eA_\LCm(x^\LCp)
	\rightarrow \int\limits_0^1\! \frac{{\dot x}^\LCm{\dot x}^\LCp}{T} + {\dot x}^\LCm eA^\LCp(\tfrac{1}{2}(x^\LCp+x^\LCm)) \;.
\ee
This is clearly nonlinear in both $x^\LCp$ and $x^\LCm$. Hence neither of the $x^\LCpm$ integrals will produce a constraining delta function and there is no reason to believe the loops or the effective action are localised. Our approach does {\it not} go through in this case. This is a positive result, because we know that the LCA does not apply to time-dependent fields $E(x^0)$.

An elegant summary of the preceding comparison is provided by following the phase-space analysis in~\cite{Dumlu:2011cc}. For the constant field, there are again two approaches leading to the same result. For the first, take $A_3 = Ex^0$, then one finds that three canonical momenta, $k_\LCperp$ and $k_3$, are conserved. (Note that $p$ is used for the canonical momentum in~\cite{Dumlu:2011cc}, whereas our $p$ is kinematic.) Integrating out the remaining component $k_0(\tau)$ then gives an effective classical action depending on $x^0(\tau)$ and constant $\{k_3,k_\LCperp\}$.  The instanton structure of this effectively 1D problem can be analysed in Euclidean space~\cite{Dumlu:2011cc}. For the second approach, take instead $A_\LCm = -\tfrac{1}{2}Ex^\LCp$ (same field, different gauge). One now finds that the three components $k_\LCperp$ and $k_\LCm$ are conserved. However, integrating out the remaining $k_\LCp(\tau)$ gives a delta functional as above, the support of which contains all the physics which would be obtained from e.g.\ the instantons. Because the effective action is gauge invariant, even though the canonical $k_\mu$ are not, these two approaches are equivalent (see also~\cite{Nikishov:2002ja,Nikishov:2002si}). Now, if the electric field is allowed to depend nontrivially on $x^\LCp$, the second option remains available, as we saw above in equations~(\ref{GP}) and (\ref{Constraint}). If, however, the field depends nontrivially on $x^0 = (x^\LCp+x^\LCm)/2$, only the first option is available, since beyond the constant field case the action cannot be made linear in any single variable, just as in (\ref{nonlinear}), and there is no way to generate a `localising' delta functional as in~(\ref{Constraint}).

\subsection{Extension to other fields}
There is a wider class of fields for which the effective action is supported only on zero-modes, and in some cases the remaining $x^\LCperp$-integrals can still be performed. Had we included a potential $A_\LCp(x^\LCp,x^\LCperp)$, the corresponding term $\dot{x}^\LCp A_\LCp$ in the action would have vanished after obtaining the delta-functional. Hence the presence of the transverse fields
\be\begin{split}\label{PVigt}
	E^i(x^\LCp,x^\LCperp) &= -\partial_i A_\LCp(x^\LCp,x^\LCperp) \;, \quad i,j\in\{1,2\}  \\
	B^i(x^\LCp,x^\LCperp) &= \epsilon^{ij} E^j(x^\LCp,x^\LCperp) \;,
\end{split}
\ee
leads to the same effective action as found above; this is natural since all such fields (a class which includes plane waves) have vanishing Schwinger invariants $F\tilde{F}=FF = 0$ and are orthogonal to the longitudinal electric field above. 

Including instead a potential $A_\LCperp(x^\LCp,x^\LCperp)$ we can describe, in addition to (\ref{PVigt}), a longitudinal magnetic field
\be
	B^3(x^\LCp,x^\LCperp) = \partial_2 A_1(x^\LCp,x^\LCperp) - \partial_1 A_2(x^\LCp,x^\LCperp) \;.
\ee
The $x^\LCm$-integrals can be performed as before, leading to localisation in $x^\LCp$. If both $E^3$ and $B^3$ depend only on $x^\LCp$, the effective action is again given exactly by the LCA applied to the known expressions for parallel, constant, $E$ and $B$~\cite{Nikishov:1969tt,Bunkin,Popov}.

\subsection{Zero-modes and the trivial vacuum}\label{trivial}
\begin{figure}[tb!!!]
	\centering\includegraphics[width=0.8\columnwidth]{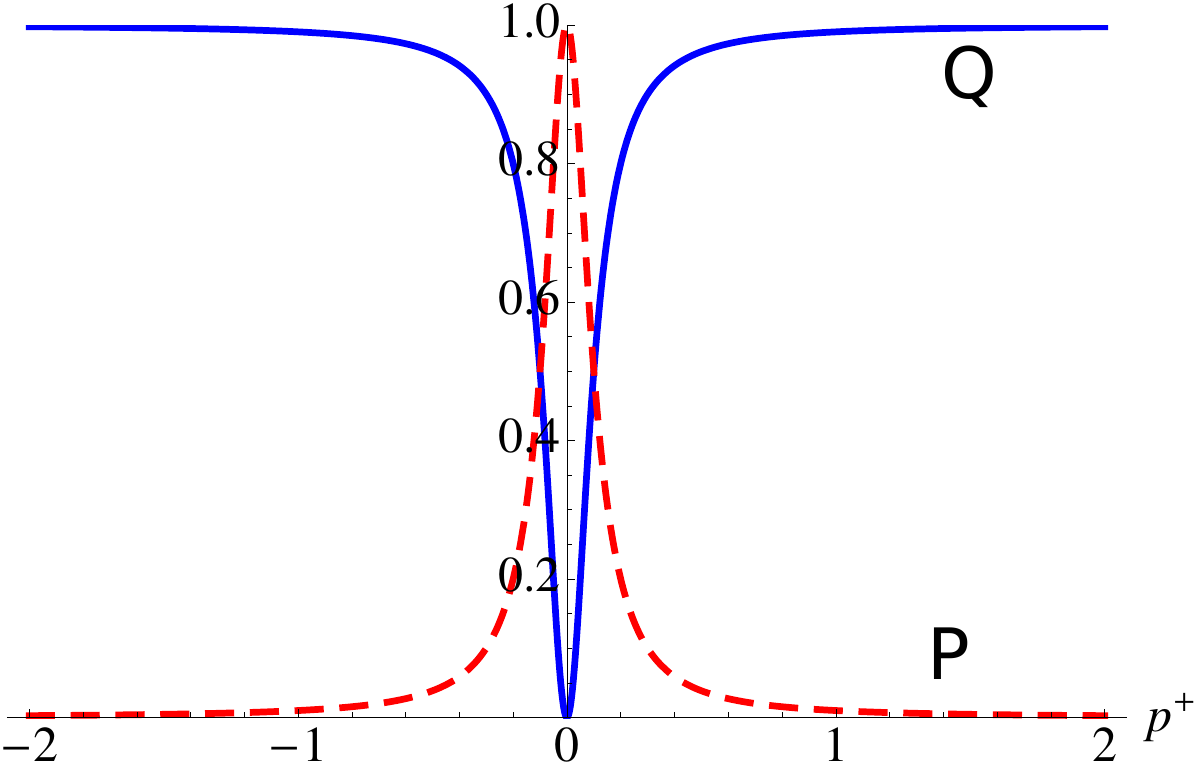}
	\caption{\label{FIG:PROJ} The zero-mode projector $\mathscr{P}$ (red/dashed curve) and non-zero-mode projector $\mathscr{Q}$ (blue/solid curve).}
\end{figure}
One of the great simplifications encountered in lightfront quantisation is that the front form vacuum is trivial up to zero-modes~\cite{Heinzl:1991vd,Ji:1995ft,Ji,Brodsky:1997de,Heinzl:2000ht}, and our calculation provides a rather explicit example of this, as we now demonstrate. We will return to the beginning of the calculation, project out the zero-modes, and see what happens.

Following~\cite{Heinzl:1991vd} we write $1=\mathscr{P} + \mathscr{Q}$ in which $\mathscr{P}$ is the zero-mode (or vacuum) projector, and $\mathscr{Q} = 1- \mathscr{P}$ is the non-zero-mode (or particle) projector. These are, in ordinary momentum space, distributions which can be constructed from nascent delta functions,
\be
	\mathscr{P} = \frac{\delta_\epsilon(p^\LCp)}{\delta_\epsilon(0)} \;, \qquad \mathscr{Q} = 1- \frac{\delta_\epsilon(p^\LCp)}{\delta_\epsilon(0)} \;,
\ee
with a standard representation being, see Fig.~\ref{FIG:PROJ},
\be
	\mathscr{P} = \frac{\epsilon^2}{{p^\LCp}^2+\epsilon^2} \;, \quad\mathscr{Q} = \frac{{p^\LCp}^2}{{p^\LCp}^2 + \epsilon^2} \;.
\ee
Since zero-modes correspond to vanishing kinematic longitudinal momentum, the natural functional equivalent of the non-zero-mode projector $\mathscr{Q}$ is
\be\label{FProj}
	\mathscr{Q} \to 1 - \frac{\delta_\epsilon[{\dot x}^\LCp]}{\delta_\epsilon[0]} \;.
\ee
(See also Appendix~\ref{APPA}.) Inserting this projector into the effective action (\ref{Gamma}) and repeating our earlier calculation, we see that the $x^\LCm$ integral can still be performed to obtain the delta functional (\ref{Constraint}). However, this now multiples the projector (\ref{FProj}), which vanishes when $\dot{x}^\LCp\equiv y^\LCp=0$, killing the whole path integral. So, when we omit the zero-modes, not only does the imaginary part of the effective action vanish (no pair production) but the {\it whole} effective action vanishes, $\Gamma\to 0$, which is the statement that the vacuum is trivial~\cite{Heinzl:2003jy}. In particular, the vacuum to vacuum transition amplitude becomes
\be
	_\text{out}\bracket{0}{0}_\text{in} = e^{i\Gamma} \to 1 \;,
\ee
consistent with a trivial vacuum, and there is no Schwinger pair production. This is recovered, though, once the zero-modes are included. A closely related result was given in~\cite{Ji:1995ft}, which considered the axial anomaly in lightfront quantisation of the massless Schwinger model. In that calculation, a manifestly gauge invariant regularisation of the chiral charge operators lead to the appearance of the gauge field zero-mode; it was then found that the rate of chiral charge production was proportional to just this zero-mode term, demonstrating the essential role of the zero-modes in pair production.

\section{Conclusions}\label{SECT:C}
The worldline approach to pair production shows that the effective action in longitudinal electric fields $E^3(x^\LCp)$ is a sum over loops constrained to lie in hypersurfaces of constant~$x^\LCp$. The localisation of these loops is the reason why the locally constant approximation is exact for the chosen class of fields. If the electric field instead depends nontrivially on $x^0$, no such localisation occurs, and indeed it is known that the locally constant approximation is not exact in that case. Indeed, it was shown in~\cite{Dunne:2005sx} that temporal field inhomogeneities tend to enhance pair production relative to the LCA, while spatial field inhomogeneities tend to suppress it. Here we have seen, following~\cite{Tomaras:2001vs,Tomaras:2000ag,Fried:2001ur}, that the `in-between' case of lightlike inhomogeneities agrees exactly with the LCA.

Both the real and imaginary parts of the effective action are supported on lightfront zero-modes. Without these, the effective action vanishes and the vacuum persistence amplitude becomes unity. This is consistent with the vacuum being trivial up to zero-modes.

It  would be interesting to compare in detail our Minkowski space calculation with the corresponding calculation in Euclidean space, and to investigate the relation to real and complex worldline instantons~\cite{Dumlu:2011cc}, as well as to the instanton calculations of~\cite{Kim:2000un,Kim:2003qp}.

\acknowledgments

A.I.\ thanks Gerald Dunne, Tom Heinzl and Christian Schubert for extremely useful discussions, and Greger Torgrimsson for a careful reading of the manuscript. A.~I.\ is supported by the {\it Swedish Research Council},  contract 2011-4221.
\appendix

\section{Mode expansions}\label{APPA}
We follow here~\cite{Polyakov:1987ez,Mansfield:1990tu}, which can be consulted for a comprehensive discussion of worldline and worldsheet methods with an emphasis on reparameterisation and conformal invariance.

The co-ordinates $x^\mu$ each have a mode expansion
\be\label{A1}
	x(\xi) = \frac{a_0}{\sqrt{T}} + \sqrt{\frac{2}{T}} \sum\limits_{n=1}^\infty a_n \cos 2n\pi \xi + b_n \sin {2n\pi\xi} \;,
\ee
and measure
\be
	\mathcal{D}x = \ud a_0\prod\limits_{n=1}\ud a_n \ud b_n \;,
\ee
as defined by the boundary conditions and the invariant inner product
\be
	(x,y) = \int\limits_0^1\!\ud\xi T\ x(\xi) y(\xi) \;.
\ee
Noting that the average position along the path is $x_c:=a_0/\sqrt{T}$, we write
\be
	x(\xi) = x_c + y(\xi) \;,
\ee
in which, from (\ref{A1}), $y(\xi)$ is orthogonal to $x_c$, i.e.\ oscillatory and non-constant. The measure may be rewritten
\be
	\mathcal{D}x  = \sqrt{T}\ud x_c \mathcal{D}y \quad \text{with} \quad  \mathcal{D}y := \prod\limits_{n=1}\ud a_n \ud b_n  \;.
\ee
Finally, the explicit form of the non-zero-mode projector introduced in Sect.~\ref{trivial} can now be seen to be
\be\label{FProj2}
	\mathscr{Q} = 1-\prod\limits_{n=1}^\infty \frac{\delta_\epsilon(a_n^\LCp)}{\delta_\epsilon(0)}\frac{\delta_\epsilon(b_n^\LCp)}{\delta_\epsilon(0)} \;.
\ee

\section{Fields which change sign}\label{SECT:TECKEN}
In the text, and as in~\cite{Tomaras:2001vs,Tomaras:2000ag}, we considered fields $E^3>0$. The arguments extend immediately to fields with $E^3<0$. If the field can change sign, though, things are not so clear. We can however give a rather formal argument for why (\ref{Constraint}) has only trivial solutions even in this case. Consider again (\ref{Constraint1-5}), which tells us that $\dot y$ is continuous (since $y$ and $a'$ are) and obeys the same boundary conditions as~$y$. Rolle's theorem then implies that there exists $\xi_*$ with ${\dot y}(\xi_*)=0$~\cite{Rolle}. Setting $\xi_0=\xi_*$ in (\ref{Constraint2}) gives $\dot{y}(\xi)\equiv 0$, so we again find $y(\xi)= 0$ as the only solution.

\end{document}